# An Impact Model of AI on the Principles of Justice: Encompassing the Autonomous Levels of AI Legal Reasoning

**Dr. Lance B. Eliot**

Chief AI Scientist, Techbruim; Fellow, CodeX: Stanford Center for Legal Informatics
Stanford, California, USA

## Abstract

Efforts furthering the advancement of Artificial Intelligence (AI) will increasingly encompass AI Legal Reasoning (AILR) as a crucial element in the practice of law. It is argued in this research paper that the infusion of AI into existing and future legal activities and the judicial structure needs to be undertaken by mindfully observing an alignment with the core principles of justice. As such, the adoption of AI has a profound twofold possibility of either usurping the principles of justice, doing so in a Dystopian manner, and yet also capable to bolster the principles of justice, doing so in a Utopian way. By examining the principles of justice across the Levels of Autonomy (LoA) of AI Legal Reasoning, the case is made that there is an ongoing tension underlying the efforts to develop and deploy AI that can demonstrably determine the impacts and sway upon each core principle of justice and the collective set.

## 1 Background and Principles of Justice

Efforts toward the advancement of Artificial Intelligence (AI) will increasingly encompass AI Legal Reasoning (AILR) as a crucial element in the practice of law [1] [7] [17] [31] [45]. It is argued in this research paper that the infusion of AI into existing and future legal activities and the judicial structure needs to be undertaken by mindfully observing an alignment with the core principles of justice.

This research paper examines the nature of the core principles of justice in Section 1. In Section 2, a framework for the Levels of Autonomy (LoA) of AI Legal Reasoning (AILR) is depicted. Then, in Section 3, the core principles of justice are aligned across the LoA AILR to showcase an anticipated evolution, along with identifying corresponding outcome facets. Section 4 provides additional considerations and proffers insights for conducting further research on these matters.

The sections in this paper are:

- Section 1: Background and Principles of Justice
- Section 2: Autonomous Levels of AI Legal Reasoning
- Section 3: Grids and Analyses of Justice Principles and LoA AILR
- Section 4: Additional Considerations and Future Research

One important assertion in this discussion is that a multi-faceted perspective should be undertaken when considering how AI will shape or reshape the instantiation of the principles of justice [7] [9] [29]. A commonly assumed false dichotomy is that the adoption of AI into the practice of law will be exclusively Dystopian or exclusively Utopian, meanwhile, it is argued herein that either possibility can arise, doing so amid each distinct principle of justice, additionally collectively so too, and that it is incumbent upon the developers and adopters of AI in the law to observe and be attune to which direction their efforts are converging [19] [38] [42].

As such, the adoption of AI has a profound twofold possibility of either usurping the principles of justice, doing so in a Dystopian manner, and yet also capable to bolster the principles of justice, doing so in a Utopian way. By examining the principles of justice across the Levels of Autonomy (LoA) of AI Legal Reasoning, the case is made that there is an ongoing tension underlying the efforts to develop and deploy AI that can demonstrably determine how each core principle of justice will be swayed.

## 1.1 identifying the Principles of Justice

Research by Susskind [44] in his book entitled "Online Courts and the Future of Justice" lays out an extensively established foundation that justice can be generally cast as consisting of seven core principles. Based on those key principles, his primary focus in the book entails an expounded argument that the emergence of online courts will both preserve justice and enhance justice, doing so via the prudent utilization of virtual hearings, plus asynchronous online judging, etc. He notably forewarns that it is not a foregone conclusion that such benefits will arise and that it will require sensible, determined, and systemic multi-generational adaptations to reach those aspirational goals.

For this research paper, the same set of core principles of justice will be utilized. This makes sense to do so herein in that the groundwork supporting the contention that those principles are indeed a bona fide and sufficient set of core principles has already been robustly well-established and therefore can be readily leveraged in a building blocks fashion accordingly, effectively and efficiently so (rather than trying to reinvent the wheel, as it were).

The emphasis herein will be to apply the emergence of AI into the practice of law amidst the core principles of justice and thus is a separate and distinct usage and analysis associated with the principles of justice in comparison to the work by Susskind that focused on the rise of online courts. Note that Susskind also identifies the significant role that AI will undoubtedly ultimately play: "In contemplating the second generation of online courts, it would be hard to ignore the recent upsurge of interest in artificial intelligence (AI) for lawyers and judges." It is that same observation of the arising spark in attention toward AI in the practice of the law that this research paper dovetails into. Similar to Susskind's argument that online courts will not axiomatically enable the principles of justice, and indeed might deter or undermine them, the same case is made herein that AI will variously have such results upon the principles of justice and that no foregone conclusion can conclusively be otherwise decreed or assumed.

The seven indicated core principles of justice consist of:

- Substantive Justice
- Procedural Justice
- Open Justice
- Distributive Justice
- Proportionate Justice
- Enforceable Justice
- Sustainable Justice

Note that the seven principles are not numbered and nor otherwise indicated as being prioritized or ranked in any particular order. They are all equally crucial. Imagine a three-legged stool that falls apart when any of the legs is missing, though in this instance envisage a seven-legged apparatus. There are tradeoffs among the principles, and it is not easy to ensure that they are each given their full and earnestly needed equal attention. Keep in mind too that existing attempts at justice are not necessarily able to live up to the ideals of the stated principles, and thus today's form of justice is undeniably at times existent of numerous shortcomings, including being too costly, taking too long, being unintelligible for many that rely upon the law, etc.

This emphasizes that today's barometer of justice is not somehow already affixed at the topmost stance [23] [31] [41]. If it were, the addition or incorporation of innovation such as the integration of AI could be argued as potentially messing with perfection, but this is not the case per se.

AI offers a chance of improving the day-to-day incurring and delivery of justice. In that same vein, if AI is improperly or inappropriately integrated, the existing justice system could be degraded, dropping from the place upon which it currently resides [8] [43].

Each of the next subsections examines each of the respective core principles of justice.

### 1.1.1 Substantive Justice

Here is an overall definition of substantive justice:

Substantive Justice: *Decisions and outcomes should be considered fair and substantive, requiring judging to be based on the laws of the land and not by whim or other divines.*

Substantive justice is about fairness, and also about the predictability of the law and what it portends [44]: "It is only fair that we are judged in accordance with whatever legislation and case law require of us. It is important in our daily activities that the law is to a great extent certain and predictable. Justice requires that judges apply law as it is, rather than what they or others think it ought to be." Also, since it is presumably possible that laws might be inherently unjust in some absolute or relativist respect, an additional criteria is that the laws as enacted and intended should be intrinsically stitched with being just [44]: "We should also insist that our justice system delivers outcomes that are themselves j*ust*."

### 1.1.2 Procedural Justice

Here is an overall definition of procedural justice:

Procedural Justice: *The process needs to be equitable and honest, independent of biases, and proffer procedures that avert the incursion of defectiveness or inconsistencies.*

Procedural justice is about the process by which justice is adjudicated, and for which if the process is skewed or malformed it can undermine and diminish the attainment of justice [44]: "A decision is considered unjust because it was handled in a manner that was in some way defective and inequitable." Within the overarching realm of procedural justice, at least two cornerstones are consisting of formal justice and natural justice. The nature of formal justice is that there should be a consistency of like cases being handled in an equivalent manner [44]: "One aspect of this concept is referred to as 'formal justice,' which is often characterized by some such phrase as 'like cases should be treated alike.'" For natural justice, it is key that a case be heard and that self-judging is to be averted [44]: "A second aspect of procedural justice is known as 'natural justice.' I am using this term in a technical sense, frequently captured in two Latin phrases: *audi alteram partem*, which requires that all litigants should be given the opportunity to state and defend their cases and *nemo index in causa sua*, which means that no-one should be a judge in his or her own case."

### 1.1.3 Open Justice

Here is an overall definition of open justice:

Open Justice: *Efforts of the courts must be transparent, open to scrutiny, accountable, and intelligible, avoiding secrecy as much as can be so reasonably achieved (realizing that at times national security, the welfare of minors, and the like can motivate some degrees of confidentiality).*

Open justice entails ensuring that processes and activities are made highly visible and shall not be unduly disguised or hidden [44]: "We object to court systems whose workings are held in private or cloaked in secrecy. We call loudly for demystification." Not all proceedings are necessarily prudent to be completely visible and therefore exceptions of an appropriate kind might need to be accounted for, but only to the extent as absolutely necessary since embracing transparency is the crux. An offshoot of the visibility aspect is that justice and all its mechanizations should be intelligible to those that are non-lawyers and thus likely unfamiliar with the nomenclature and complexities of the law [44]: "There are strong arguments in support of the view that open justice also requires any information and data findings about the courts, as well as the court proceedings themselves, to be understandable to non-lawyers."

### 1.1.4 Distributive Justice

Here is an overall definition of distributive justice:

Distributive Justice: *Each person must be given their legal due and afforded access to justice, thus driving a semblance of distributiveness to ensure that regardless of means that all can gain access.*

Distributive justice entails seeking to make access to justice feasible since otherwise, denial of access is essentially no different from an altogether lack of justice itself [44]: "Distributive justice requires that court service is accessible and intelligible to all; that access to legal and court services is a benefit that is evenly spread across society; that rights and duties are equally allocated; that the powerful and rich are subject to the same law as the less well-off and less powerful; and that the service is affordable by all regardless of their means."

### 1.1.5 Proportionate Justice

Here is an overall definition of proportionate justice:

Proportionate Justice: *Fairness ought to arise at scale, straightforward processes for straightforward issues, attempting to ensure that speediness occurs and aligns too with complexity, suitable proportionality based on the assertion that justice delayed is justice diluted.*

Proportionality is a less often enumerated component of justice and tends to be assumed as existent or deemed as unworthy of being an *essential* element of justice and can be perhaps cast as a secondary condition, but here it is viewed as equally vital as the other principles. In brief, the notion is that the energy or effort of achieving justice should be proportional to the nature or magnitude of the underlying dispute for which justice is being sought, thus [44]: "The principle of proportionality requires, first of all, that we should ensure that the cost of handling individual cases in our courts makes sense by reference to the nature and value of each dispute." And, regarding our revered adversarial approach [44]: "In a similar vein, although our system is adversarial, this should not mean that all disputes are conducted in a highly combative spirit. Unwarranted escalation of disputes, especially in smaller cases, should be discouraged."

### 1.1.6 Enforceable Justice

Here is an overall definition of enforceable justice:

Enforceable Justice: *Results need to have teeth and be seen as binding, enforcement as enabled via the coercive power of the state, correctly deprive money and property and liberty to ensure justice is served.*

Enforceable justice entails the need to ensure that justice must not be hollow, namely that if there was no means or mechanism to enforce or bind the ascertained results then there would be no consequent impact or effect per se of having sought justice. In a sense, justice would have no semblance of potency since it could not be otherwise implemented or compelled [44]: "The determination of judges are binding and can be enforced by the coercive power of the state." And without such implementations, there would seem little reason for society to avail themselves of relying upon the efforts of the judiciary: "Without enforceable justice, the law runs the risk of affording a rather weak set of protections."

### 1.1.7 Sustainable Justice

Here is an overall definition of sustainable justice:

Sustainable Justice: *Have a stable basis for the ongoing instantiation of justice, sufficient resources must be allocated to maintain and incur upkeep for continually improving the means of the courts to act, including being able to demonstrably scale to whatever volume of cases might be presented.*

Sustainable justice necessitates the somewhat abstract but very real notion that justice and all of its elements must be in existence; otherwise, if it is intermittent or known to be unreliably sustained then such justice cannot be depended upon [44]: "Courts should be safe havens; solid and reliable; anchors to which, in times of need, people and organizations can confidently tether themselves." As an aside, it can be asserted that the sustainability of justice also intertwines with the technological capabilities of those being served [44]: "It is also hard to conceive of a truly sustainable court system that is not technologically in tune with the communities that it serves."

## 2 Autonomous Levels of AI Legal Reasoning

In this section, a framework for the autonomous levels of AI Legal Reasoning is summarized and is based on the research described in detail in Eliot [20].

These autonomous levels will be portrayed in a grid that aligns with key elements of autonomy and as matched to AI Legal Reasoning. Providing this context will be useful to the later sections of this paper and will be utilized accordingly.

The autonomous levels of AI Legal Reasoning are as follows:

Level 0: No Automation for AI Legal Reasoning
Level 1: Simple Assistance Automation for AI Legal Reasoning
Level 2: Advanced Assistance Automation for AI Legal Reasoning
Level 3: Semi-Autonomous Automation for AI Legal Reasoning
Level 4: Domain Autonomous for AI Legal Reasoning
Level 5: Fully Autonomous for AI Legal Reasoning
Level 6: Superhuman Autonomous for AI Legal Reasoning

## 2.1 Details of the LoA AILR

See **Figure A-1** for an overview chart showcasing the autonomous levels of AI Legal Reasoning as via columns denoting each of the respective levels.

See **Figure A-2** for an overview chart similar to Figure A-1 which alternatively is indicative of the autonomous levels of AI Legal Reasoning via the rows as depicting the respective levels (this is simply a reformatting of Figure A-1, doing so to aid in illuminating this variant perspective, but does not introduce any new facets or alterations from the contents as already shown in Figure A-1).

### 2.1.1 Level 0: No Automation for AI Legal Reasoning

Level 0 is considered the no automation level. Legal reasoning is carried out via manual methods and principally occurs via paper-based methods.

This level is allowed some leeway in that the use of say a simple handheld calculator or perhaps the use of a fax machine could be allowed or included within this Level 0, though strictly speaking it could be said that any form whatsoever of automation is to be excluded from this level.

### 2.1.2 Level 1: Simple Assistance Automation for AI Legal Reasoning

Level 1 consists of simple assistance automation for AI legal reasoning.

Examples of this category encompassing simple automation would include the use of everyday computer-based word processing, the use of everyday computer-based spreadsheets, the access to online legal documents that are stored and retrieved electronically, and so on.

By-and-large, today's use of computers for legal activities is predominantly within Level 1. It is assumed and expected that over time, the pervasiveness of automation will continue to deepen and widen, and eventually lead to legal activities being supported and within Level 2, rather than Level 1.

### 2.1.3 Level 2: Advanced Assistance Automation for AI Legal Reasoning

Level 2 consists of advanced assistance automation for AI legal reasoning.

Examples of this notion encompassing advanced automation would include the use of query-style Natural Language Processing (NLP), Machine Learning (ML) for case predictions, and so on.

Gradually, over time, it is expected that computer-based systems for legal activities will increasingly make use of advanced automation. Law industry technology that was once at a Level 1 will likely be refined, upgraded, or expanded to include advanced capabilities, and thus be reclassified into Level 2.

### 2.1.4 Level 3: Semi-Autonomous Automation for AI Legal Reasoning

Level 3 consists of semi-autonomous automation for AI legal reasoning.

Examples of this notion encompassing semi-autonomous automation would include the use of Knowledge-Based Systems (KBS) for legal reasoning, the use of Machine Learning and Deep Learning (ML/DL) for legal reasoning, and so on.

Today, such automation tends to exist in research efforts or prototypes and pilot systems, along with some commercial legal technology that has been infusing these capabilities too.

### 2.1.5 Level 4: Domain Autonomous for AI Legal Reasoning

Level 4 consists of domain autonomous computer-based systems for AI legal reasoning.

This level reuses the conceptual notion of Operational Design Domains (ODDs) as utilized in the autonomous vehicles and self-driving cars levels of autonomy, though in this use case it is being applied to the legal domain [17] [18] [20].

Essentially, this entails any AI legal reasoning capacities that can operate autonomously, entirely so, but that is only able to do so in some limited or constrained legal domain.

## 2.1.6 Level 5: Fully Autonomous for AI Legal Reasoning

Level 5 consists of fully autonomous computer-based systems for AI legal reasoning.

In a sense, Level 5 is the superset of Level 4 in terms of encompassing all possible domains as per however so defined ultimately for Level 4. The only constraint, as it were, consists of the facet that the Level 4 and Level 5 are concerning human intelligence and the capacities thereof. This is an important emphasis due to attempting to distinguish Level 5 from Level 6 (as will be discussed in the next subsection)

It is conceivable that someday there might be a fully autonomous AI legal reasoning capability, one that encompasses all of the law in all foreseeable ways, though this is quite a tall order and remains quite aspirational without a clear cut path of how this might one day be achieved. Nonetheless, it seems to be within the extended realm of possibilities, which is worthwhile to mention in relative terms to Level 6.

## 2.1.7 Level 6: Superhuman Autonomous for AI Legal Reasoning

Level 6 consists of superhuman autonomous computer-based systems for AI legal reasoning.

In a sense, Level 6 is the entirety of Level 5 and adds something beyond that in a manner that is currently ill-defined and perhaps (some would argue) as yet unknowable. The notion is that AI might ultimately exceed human intelligence, rising to become superhuman, and if so, we do not yet have any viable indication of what that superhuman intelligence consists of and nor what kind of thinking it would somehow be able to undertake.

Whether a Level 6 is ever attainable is reliant upon whether superhuman AI is ever attainable, and thus, at this time, this stands as a placeholder for that which might never occur. In any case, having such a placeholder provides a semblance of completeness, doing so without necessarily legitimatizing that superhuman AI is going to be achieved or not. No such claim or dispute is undertaken within this framework.

## 3 Grids and Analyses of the Principles of Justice and LoA AILR

In this section, the autonomous levels of AI Legal Reasoning will be aligned with the principles of justice, forming a grid that is indicative of how AI might impact the principles and likewise how the principles can be utilized to drive the development and maturation of AI for the law. As for nomenclature, the nature of impacts involving the LoA AILR on the principles of justice is abbreviated as LoA-principles, while the impacts of the principles of justice on the LoA AILR is denoted as principles-LoA.

### 3.1 Principles of Justice and the LoA AILR

As shown in **Figure B-1**, it is useful and informative to align the seven core principles of justice with the seven levels of autonomy of AI Legal Reasoning. An explanation of the grid and its significance is discussed next.

For Level 0, Level 1, and Level 2, the grid indicates that the alignment is considered as "Traditional" in the sense that since the level of automation is conventional at those levels of LoA this ergo suggests that any impacts on or of the principles of justice consist of what we already know and anticipate. Level 3 is the first turning point as it is considered the semi-autonomous LoA and thus is expressed as "Emerging," which means that the impacts will start to become notable about the autonomy that might then emerge or exist once the Level 4 and above are achieved. Level 4 is the AILR domain autonomous level and the impacts are denoted as Phase X, which is explained in subsequent charts. Level 5 is the AILR fully autonomous level and the impacts are denoted as Phase Y, which is explained in subsequent charts. Level 6 is the AILR superhuman autonomous level and the impacts are denoted as Phase Z, which is explained in subsequent charts.

In a recap of this grid:

**Substantive Justice**
- Level 0: Traditional
- Level 1: Traditional
- Level 2: Traditional
- Level 3: Emerging
- Level 4: Phase X Impacts
- Level 5: Phase Y Impacts
- Level 6: Phase Z Impacts

**Procedural Justice**
- Level 0: Traditional
- Level 1: Traditional
- Level 2: Traditional
- Level 3: Emerging
- Level 4: Phase X Impacts
- Level 5: Phase Y Impacts
- Level 6: Phase Z Impacts

**Open Justice**
- Level 0: Traditional
- Level 1: Traditional
- Level 2: Traditional
- Level 3: Emerging
- Level 4: Phase X Impacts
- Level 5: Phase Y Impacts
- Level 6: Phase Z Impacts

**Distributive Justice**
- Level 0: Traditional
- Level 1: Traditional
- Level 2: Traditional
- Level 3: Emerging
- Level 4: Phase X Impacts
- Level 5: Phase Y Impacts
- Level 6: Phase Z Impacts

**Proportionate Justice**
- Level 0: Traditional
- Level 1: Traditional
- Level 2: Traditional
- Level 3: Emerging
- Level 4: Phase X Impacts
- Level 5: Phase Y Impacts
- Level 6: Phase Z Impacts

**Enforceable Justice**
- Level 0: Traditional
- Level 1: Traditional
- Level 2: Traditional
- Level 3: Emerging
- Level 4: Phase X Impacts
- Level 5: Phase Y Impacts
- Level 6: Phase Z Impacts

**Sustainable Justice**
- Level 0: Traditional
- Level 1: Traditional
- Level 2: Traditional
- Level 3: Emerging
- Level 4: Phase X Impacts
- Level 5: Phase Y Impacts
- Level 6: Phase Z Impacts

**Figure B-2** is akin to Figure B-1, illustrating the same grid but with the LoA along the rows and the core principles of justice indicated as the columns. This is not a new introduction of facets and instead merely a convenient means of representing the content in a flipped format for ease of reference and furtherance to the discussion.

In a recap of the grid:

**Level 0: No Automation**
- Substantive Justice: *Traditional*
- Procedural Justice: *Traditional*
- Open Justice: *Traditional*
- Distributive Justice: *Traditional*
- Proportionate Justice: *Traditional*
- Enforceable Justice: *Traditional*
- Sustainable Justice: *Traditional*

**Level 1: Simple Assistance Automation**
- Substantive Justice: *Traditional*
- Procedural Justice: *Traditional*
- Open Justice: *Traditional*
- Distributive Justice: *Traditional*
- Proportionate Justice: *Traditional*
- Enforceable Justice: *Traditional*
- Sustainable Justice: *Traditional*

**Level 2: Advanced Assistance Automation**
- Substantive Justice: *Traditional*
- Procedural Justice: *Traditional*
- Open Justice: *Traditional*
- Distributive Justice: *Traditional*

- Proportionate Justice: *Traditional*
- Enforceable Justice: *Traditional*
- Sustainable Justice: *Traditional*

**Level 3: Semi-Autonomous Automation**
- Substantive Justice: *Emerging*
- Procedural Justice: *Emerging*
- Open Justice: *Emerging*
- Distributive Justice: *Emerging*
- Proportionate Justice: *Emerging*
- Enforceable Justice: *Emerging*
- Sustainable Justice: *Emerging*

**Level 4: AILR Domain Autonomous**
- Substantive Justice: *Phase X Impacts*
- Procedural Justice: *Phase X Impacts*
- Open Justice: *Phase X Impacts*
- Distributive Justice: *Phase X Impacts*
- Proportionate Justice: *Phase X Impacts*
- Enforceable Justice: *Phase X Impacts*
- Sustainable Justice: *Phase X Impacts*

**Level 5: AILR Fully Autonomous**
- Substantive Justice: *Phase Y Impacts*
- Procedural Justice: *Phase Y Impacts*
- Open Justice: *Phase Y Impacts*
- Distributive Justice: *Phase Y Impacts*
- Proportionate Justice: *Phase Y Impacts*
- Enforceable Justice: *Phase Y Impacts*
- Sustainable Justice: *Phase Y Impacts*

**Level 6: AILR Superhuman Autonomous**
- Substantive Justice: *Phase Z Impacts*
- Procedural Justice: *Phase Z Impacts*
- Open Justice: *Phase Z Impacts*
- Distributive Justice: *Phase Z Impacts*
- Proportionate Justice: *Phase Z Impacts*
- Enforceable Justice: *Phase Z Impacts*
- Sustainable Justice: *Phase Z Impacts*

**3.2 AI Infusion and Potential Outcomes**

Shown in **Figure B-3** is an indication of the seven core principles of justice and the notable aspect that the AI infusion can be construed in a twofold manner, consisting of outcomes that give rise to Utopian results and also consisting of outcomes that give rise to Dystopian results.

These are not to be considered as mutually exclusive of each other, in the sense that the results can vary, both by the distinct principles of justice, each individually so, and also collectively across the entire set. It is postulated that the results will be differentiated across the LoA AILR in the sense that there is a measured difference between the Level 4, Level 5, and Level 6, thus each of those level distinctions is given a phasing, respectively indicated as Phase X, Phase Y, and Phase Z (as will be explained further in the next subsection).

Note that this is not a prescriptive indication and thus there is no suggestion, implication, or proclamation that there will be Utopian results, nor that there will be Dystopian results, and that instead this is providing a means of being able to both *reactively* assess the results of AI infusion and can also be *proactively* used to drive the directional nature of an AI infusion into the law.

**3.3 Phase X, Phase Y, Phase Z**

Shown in **Figure B-4** is an indication of the seven core principles and the respective Phase X, Phase Y, and Phase Z indications. Within each Level, the respective phase showcases the possibility of a Utopian outcome and a Dystopian outcome, per each of the respective core principles of justice. Note that the wording is evocative of being more so or less of the characteristic stated, such as the word "fairer" to indicate that the outcome would potentially be heightened in fairness, while the word "unfairer" to indicate that the outcome would potentially be heightened in unfairness.

*Phase X is designated as the impacts at Level 4 and consists of:*

Phase X: Level 4
Substantive Justice
Utopian: Fairer Decisions (law domains)
Dystopian: Unfairer Decisions (law domains)

Phase X: Level 4
Procedural Justice
Utopian: Fairer Processes (law domains)
Dystopian: Unfairer Processes (law domains)

Phase X: Level 4
Open Justice
Utopian: Greater Transparency (law domains)
Dystopian: Lessened Transparency (law domains)

Phase X: Level 4
Distributive Justice
Utopian: Expanded Access (law domains)
Dystopian: Reduced Access (law domains)

Phase X: Level 4
Proportionate Justice
Utopian: More Balanced (law domains)
Dystopian: Less Balanced (law domains)

Phase X: Level 4
Enforceable Justice
Utopian: Better Enforcement (law domains)
Dystopian: Worse Enforcement (law domains)

Phase X: Level 4
Sustainable Justice
Utopian: Greater Stability (law domains)
Dystopian: Lessened Stability (law domains)

*Phase Y is designated as the impacts at Level 5 and consists of:*

Phase Y: Level 5
Substantive Justice
Utopian: Fairer Decisions (all of law)
Dystopian: Unfairer Decisions (all of law)

Phase Y: Level 5
Procedural Justice
Utopian: Fairer Processes (all of law)
Dystopian: Unfairer Processes (all of law)

Phase Y: Level 5
Open Justice
Utopian: Greater Transparency (all of law)
Dystopian: Lessened Transparency (all of law)

Phase Y: Level 5
Distributive Justice
Utopian: Expanded Access (all of law)
Dystopian: Reduced Access (all of law)

Phase Y: Level 5
Proportionate Justice
Utopian: More Balanced (all of law)
Dystopian: Less Balanced (all of law)

Phase Y: Level 5
Enforceable Justice
Utopian: Better Enforcement (all of law)
Dystopian: Worse Enforcement (all of law)

Phase Y: Level 5
Sustainable Justice
Utopian: Greater Stability (all of law)
Dystopian: Lessened Stability (all of law)

*Phase Z is designated as the impacts at Level 6 and consists of:*

Phase Z: Level 6
Substantive Justice
Utopian: Ultra-Fair Decisions
Dystopian: Ultra-Unfair Decisions

Phase Z: Level 6
Procedural Justice
Utopian: Ultra-Fair Processes
Dystopian: Ultra-Unfair Processes

Phase Z: Level 6
Open Justice
Utopian: Ultimate Transparency
Dystopian: Utmost Opaqueness

Phase Z: Level 6
Distributive Justice
Utopian: Totality of Access
Dystopian: Utter Denial of Access

Phase Z: Level 6
Proportionate Justice
Utopian: Perfectly Balanced
Dystopian: Wholly Unbalanced

Phase Z: Level 6
Enforceable Justice
Utopian: Idea Enforcement
Dystopian: Horrendous Enforcement

Phase Z: Level 6
Sustainable Justice
Utopian: Completely Sustainable
Dystopian: Entirely Unsustainable

## 3.4 Examples of Impacts Analysis

When considering how AI infusion will potentially impact justice, the grids can be substantively conducive in at least two ways: (a) conducting an impact analysis anew, and (b) used when evaluating an existing impact analysis that is being presented by a given research study.

For any such analyses, questions to be asked and appropriately addressed include:

- Which of the core principles of justice is being included?

- Which of the core principles of justice is not being included and how might that omission alter the otherwise predicted impacts?

- Is the tension and balancing among the core principles of justice being encompassed?

- At what level of AI autonomy is the focus being undertaken?

- To what degree and nature do the AI impact a specific principle of justice?

- How does the AI impact across the core set of principles of justice?

- Is a predominantly Utopian perspective being assumed?

- Is a predominantly Dystopian perspective being assumed?

- Does the analysis consider the separate impact by core principle rather than as a monolith?

- Does the analysis consider the collective impacts encompassing the set of principles?

- And so on.

Suppose for example that a research study has asserted that AI will lead to fairer decisions by the justice system. Consider using the grids to examine and assess the conclusion reached by this exemplar.

As viewed within the context of the grids, this might be equated with an assertion that entails Substantive Justice (due to invoking the notion of fairness in decisions) and posits a Utopian leaning impact (due to avowing that decisions will be fairer, rather than just as fair or perhaps even turning toward unfair). If so, there should presumably be a rational or justification provided in the research study as to not only why the Utopian leaning is warranted as an outcome, but would also need to ascertain why the Dystopian is unlikely to occur in lieu of the Utopian claim.

Meanwhile, this still leaves unstated by the research study as to what level of AI is being assumed, such as whether this is at a level below the autonomous levels of Level 4, Level 5, or Level 6, or those respective levels. This exposes a weakness in the impacts being alleged since the magnitude and scope of the AI infusion is either omitted or otherwise not explicitly addressed. Furthermore, focusing on just one principle, in this example the Substantive Justice principle, entirely undercuts the dependency nature of the core principles among each other. If there are apparently going to be fairer decisions in the realm of Substantive Justice, this might come at the cost of perhaps unfairer processes (i.e., worsened Procedural Justice) or at the cost of less Open Justice (lessened transparency), and so on. By consulting the grid, it becomes readily feasible to directly seek to discover whether a research study has covered the range of principles of justice. By the happenstance or oversight of not considering the full range, it is conceivable that a presumed optimization at one principle could seemingly undercut the performance on one or more of the other principles.

These grids then provide a tool or model for the pursuit of research on the impacts of AI on the principles of justice and the impacts of the principles on the development and deployment of AI.

## 4 Additional Considerations and Future Research

As earlier stated, efforts toward the advancement of Artificial Intelligence (AI) will increasingly encompass AI Legal Reasoning (AILR) as a crucial element in the practice of law. It has been argued in this research paper that the infusion of AI into existing and future legal activities and the judicial structure needs to be undertaken by mindfully observing an alignment with the core principles of justice. The adoption of AI has a profound twofold possibility of either usurping the principles of justice, doing so in a Dystopian manner, and yet also capable to bolster the principles of justice, doing so in a Utopian way.

Any research examining the principles of justice and AI must strive to be as complete and cohesive as feasible, for which the grids provided in this paper are a proffered tool(s) or model(s) to be so productively utilized. Future research for evolving these tools or models provided is highly recommended. This might consist of the following types of studies:

- Conduct case studies using the tools or models
- Analyze existing research studies using the tools or models
- Craft new research that encompasses the tools or models
- Propose extensions to the tools or models
- Other

By considering the principles of justice across the Levels of Autonomy (LoA) of AI Legal Reasoning, and then addressing the singular facets of each principle, along with the collective set, research in this realm will assuredly be more robust. Also, undertaking such analyses will allow for greater comparison of such research studies, readily revealing which assumptions are being made, including any omissions or oversights, and aid in bolstering the state of research on these matters.

**About the Author**

Dr. Lance Eliot is the Chief AI Scientist at Techbrium Inc. and a Stanford Fellow at Stanford University in the CodeX: Center for Legal Informatics. He previously was a professor at the University of Southern California (USC) where he headed a multi-disciplinary and pioneering AI research lab. Dr. Eliot is globally recognized for his expertise in AI and is the author of highly ranked AI books and columns.

**Figure A-1**

| AI & Law: Levels of Autonomy For AI Legal Reasoning (AILR) | | | | |
|:---:|:---:|:---:|:---:|:---:|
| **Level** | **Descriptor** | **Examples** | **Automation** | **Status** |
| **0** | No Automation | Manual, paper-based (no automation) | None | De Facto - In Use |
| **1** | Simple Assistance Automation | Word Processing, XLS, online legal docs, etc. | Legal Assist | Widely In Use |
| **2** | Advanced Assistance Automation | Query-style NLP, ML for case prediction, etc. | Legal Assist | Some In Use |
| **3** | Semi-Autonomous Automation | KBS & ML/DL for legal reasoning & analysis, etc. | Legal Assist | Primarily Prototypes & Research Based |
| **4** | AILR Domain Autonomous | Versed only in a specific legal domain | Legal Advisor (law fluent) | None As Yet |
| **5** | AILR Fully Autonomous | Versatile within and across all legal domains | Legal Advisor (law fluent) | None As Yet |
| **6** | AILR Superhuman Autonomous | Exceeds human-based legal reasoning | Supra Legal Advisor | Indeterminate |

*Figure 1: AI & Law - Autonomous Levels by Rows*          *Source Author: Dr. Lance B. Eliot*

V1.3

**Figure A-2**

## AI & Law: Levels of Autonomy For AI Legal Reasoning (AILR)

| | Level 0 | Level 1 | Level 2 | Level 3 | Level 4 | Level 5 | Level 6 |
|---|---|---|---|---|---|---|---|
| **Descriptor** | No Automation | Simple Assistance Automation | Advanced Assistance Automation | Semi-Autonomous Automation | AILR Domain Autonomous | AILR Fully Autonomous | AILR Superhuman Autonomous |
| **Examples** | Manual, paper-based (no automation) | Word Processing, XLS, online legal docs, etc. | Query-style NLP, ML for case prediction, etc. | KBS & ML/DL for legal reasoning & analysis, etc. | Versed only in a specific legal domain | Versatile within and across all legal domains | Exceeds human-based legal reasoning |
| **Automation** | None | Legal Assist | Legal Assist | Legal Assist | Legal Advisor (law fluent) | Legal Advisor (law fluent) | Supra Legal Advisor |
| **Status** | De Facto – In Use | Widely In Use | Some In Use | Primarily Prototypes & Research-based | None As Yet | None As Yet | Indeterminate |

*Figure 2: AI & Law - Autonomous Levels by Columns*          *Source Author: Dr. Lance B. Eliot*

V1.3

**Figure B-1**

## Principles of Justice and Autonomous Levels of AI Legal Reasoning (AILR)

| | Level 0 | Level 1 | Level 2 | Level 3 | Level 4 | Level 5 | Level 6 |
|---|---|---|---|---|---|---|---|
| Descriptor | No Automation | Simple Assistance Automation | Advanced Assistance Automation | Semi-Autonomous Automation | AILR Domain Autonomous | AILR Fully Autonomous | AILR Superhuman Autonomous |
| **Substantive Justice** | *Traditional* | *Traditional* | *Traditional* | Emerging | Phase X Impacts | Phase Y Impacts | Phase Z Impacts |
| **Procedural Justice** | *Traditional* | *Traditional* | *Traditional* | Emerging | Phase X Impacts | Phase Y Impacts | Phase Z Impacts |
| **Open Justice** | Traditional | Traditional | Traditional | Emerging | Phase X Impacts | Phase Y Impacts | Phase Z Impacts |
| **Distributive Justice** | *Traditional* | *Traditional* | *Traditional* | Emerging | Phase X Impacts | Phase Y Impacts | Phase Z Impacts |
| **Proportionate Justice** | Traditional | Traditional | Traditional | Emerging | Phase X Impacts | Phase Y Impacts | Phase Z Impacts |
| **Enforceable Justice** | *Traditional* | *Traditional* | *Traditional* | Emerging | Phase X Impacts | Phase Y Impacts | Phase Z Impacts |
| **Sustainable Justice** | *Traditional* | *Traditional* | *Traditional* | Emerging | Phase X Impacts | Phase Y Impacts | Phase Z Impacts |

*Figure 1: AI & Law – Principles of Justice and LoA AILR by Columns*          *Source Author: Dr. Lance B. Eliot*     V1.3

**Figure B-2**

## Principles of Justice and Levels of Autonomy For AI Legal Reasoning (AILR)

| Level | Descriptor | Substantive Justice | Procedural Justice | Open Justice | Distributive Justice | Proportionate Justice | Sustainable Justice |
|-------|-----------|-----------|-----------|-----------|-----------|-----------|-----------|
| 0 | No Automation | *Traditional* | *Traditional* | *Traditional* | *Traditional* | *Traditional* | *Traditional* |
| 1 | Simple Assistance Automation | *Traditional* | *Traditional* | *Traditional* | *Traditional* | *Traditional* | *Traditional* |
| 2 | Advanced Assistance Automation | *Traditional* | *Traditional* | *Traditional* | *Traditional* | *Traditional* | *Traditional* |
| 3 | Semi-Autonomous Automation | Emerging | Emerging | Emerging | Emerging | Emerging | Emerging |
| 4 | AILR Domain Autonomous | Phase X Impacts | Phase X Impacts | Phase X Impacts | Phase X Impacts | Phase X Impacts | Phase X Impacts |
| 5 | AILR Fully Autonomous | Phase Y Impacts | Phase Y Impacts | Phase Y Impacts | Phase Y Impacts | Phase Y Impacts | Phase Y Impacts |
| 6 | AILR Superhuman Autonomous | Phase Z Impacts | Phase Z Impacts | Phase Z Impacts | Phase Z Impacts | Phase Z Impacts | Phase Z Impacts |

*Figure 2: AI & Law – Principles of Justice and LoA AILR by Rows*     *Source Author: Dr. Lance B. Eliot*  V1.3

**Figure B-3**

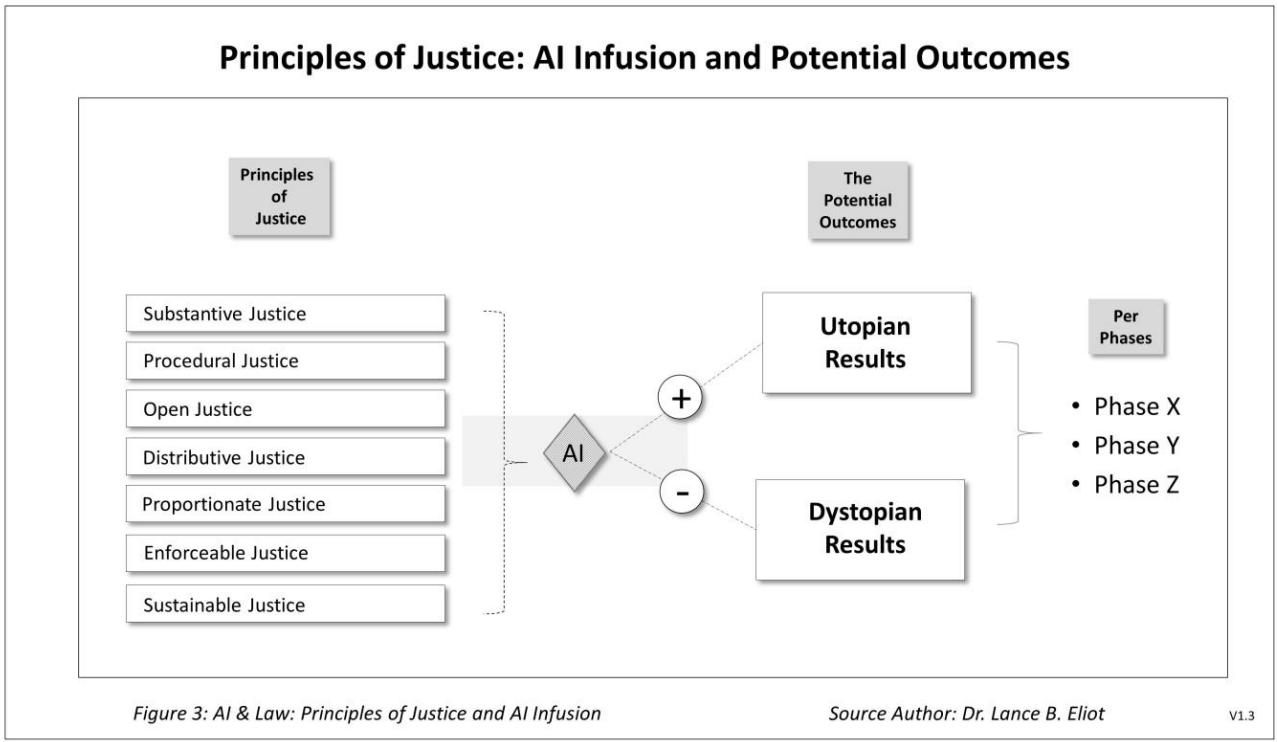

**Principles of Justice: AI Infusion and Potential Outcomes**

Principles of Justice

The Potential Outcomes

Substantive Justice
Procedural Justice
Open Justice
Distributive Justice
Proportionate Justice
Enforceable Justice
Sustainable Justice

AI

+

−

Utopian Results

Dystopian Results

Per Phases

• Phase X
• Phase Y
• Phase Z

*Figure 3: AI & Law: Principles of Justice and AI Infusion*     *Source Author: Dr. Lance B. Eliot*     V1.3

**Figure B-4**

## Principles of Justice Outcomes and LoA of AI Legal Reasoning (AILR)

| | Level 4 | Level 4 | Level 5 | Level 5 | Level 6 | Level 6 |
|---|---|---|---|---|---|---|
| | *Phase X* | | *Phase Y* | | *Phase Z* | |
| Descriptor | *Utopian* | *Dystopian* | *Utopian* | *Dystopian* | *Utopian* | *Dystopian* |
| **Substantive Justice** | Fairer Decisions (law domains) | Unfairer Decisions (law domains) | Fairer Decisions (all of law) | Unfairer Decisions (all of law) | Ultra-Fair Decisions | Ultra-Unfair Decisions |
| **Procedural Justice** | Fairer Processes (law domains) | Unfairer Processes (law domains) | Fairer Processes (all of law) | Unfairer Processes (all of law) | Ultra-Fair Processes | Ultra-Unfair Processes |
| **Open Justice** | Greater Transparency (law domains) | Lessened Transparency (law domains) | Greater Transparency (all of law) | Lessened Transparency (all of law) | Ultimate Transparency | Utmost Opaqueness |
| **Distributive Justice** | Expanded Access (law domains) | Reduced Access (law domains) | Expanded Access (all of law) | Reduced Access (all of law) | Totality of Access | Utter Denial of Access |
| **Proportionate Justice** | More Balanced (law domains) | Less Balanced (law domains) | More Balanced (all of law) | Less Balanced (all of law) | Perfectly Balanced | Wholly Unbalanced |
| **Enforceable Justice** | Better Enforcement (law domains) | Worse Enforcement (law domains) | Better Enforcement (all of law) | Worse Enforcement (all of law) | Ideal Enforcement | Horrendous Enforcement |
| **Sustainable Justice** | Greater Stability (law domains) | Lessened Stability (law domains) | Greater Stability (all of law) | Lessened Stability (all of law) | Completely Sustainable | Entirely Unsustainable |

*Strawman Variant*

*Figure 4: AI & Law – Principles of Justice Outcomes and LoA AILR by Columns*     Source Author: Dr. Lance B. Eliot     V1.3